# MULTIPLEX VISIBILITY GRAPHS AS A COMPLEMENTARY TOOL FOR DESCRIBING THE RELATION BETWEEN GROUND LEVEL $O_3$ AND $NO_2$

Carmona-Cabezas Rafael[1,*], Gómez-Gómez Javier[1], Ariza-Villaverde Ana B.[1], Gutiérrez de Ravé Eduardo[1], Jiménez-Hornero Francisco J.[1]

[1] Complex Geometry, Patterns and Scaling in Natural and Human Phenomena (GEPENA) Research Group, University of Cordoba, Gregor Mendel Building (3rd floor), Campus Rabanales, 14071 Cordoba, Spain

* Corresponding author. e-mail: f12carcr@uco.es

## DECLARATION OF INTERESTS

Declarations of interest: none

MVG: Multiplex Visibility Graph; VG: Visibility Graph

ABSTRACT

The usage of multilayer complex networks for the analysis of correlations among environmental variables (such as $O_3$ and $NO_2$ concentrations from the photochemical smog) is investigated in this work. The mentioned technique is called *Multiplex Visibility Graphs (MVG).* By performing the joint analysis of those layers, the parameters named Average Edge Overlap and Interlayer Mutual Information are extracted, which accounts for the microscopical time coherence and the correlation between the time series behavior, respectively.

These parameters point to the possibility of using them independently to describe the correlation between atmospheric pollutants (which could be extended to environmental time series). More precisely the first one of them is considered to be a potential new approach to determine the time required for the correlation of $NO_2$ and $O_3$ to be observed, since it is obtained from the correlation of the pollutants at the smallest time scale. As for the second one, it has been checked that the proposed technique can be used to describe the variation of the correlation between the two gases along the seasons. In short, MVGs parameters are introduced and results show that they could be potentially used in a future for correlation studies, supplementing already existing techniques.

1. <u>INTRODUCTION</u>

In the last years, many studies have been conducted to give some light on formation and dynamics of ground-level ozone. The important of these analyses lie on the fact that it is one of the main photochemical oxidants (due to its abundance) and it can lead to serious damage for human health and harvest for high concentrations (Doherty et al., 2009). According to Miao (Miao et al., 2017), its impact can be quantified in losses of billions of dollars from the economical point of view.

The formation and destruction of this secondary pollutant is known to be governed by photochemical and nonlinear processes (Graedel and Crutzen, 1993; Trainer et al., 2000) that depend highly on meteorological conditions such as temperature, wind direction and mainly solar radiation (Trainer et al., 2000). In addition to all that, $O_3$ concentration also depends on the behavior of its chemical precursors. These are mainly nitrogen oxides (amongst them, $NO_2$ is studied here) and volatile organic compounds produced from the urban and industrial activity (Sillman, 1999). It is because of all those factors that the analysis of the temporal evolution of ozone is a very complex task. Hence, some questions remain open such as the influence of the working time scale on these studies and the relevance of $NO_2$ as precursor depending on the season of the year.

A very recent method (called *Multiplex Visibility Graph*, MVG) to analyze nonlinear multivariate timeseries (Lacasa et al., 2015) is used in this work to see the relation between $O_3$ and its precursor $NO_2$ to answer the questions posed above. This method consists basically on turning the time series into complex networks and then forming multilayered structures that can be analyzed afterward, thanks to the last advances in this field. MVGs have shown to be useful for several applications already, from economics to neurology (Bianchi et al., 2017; Lacasa et al., 2015; Sannino et al., 2017). Those works have used last developments in multilayer networks (Bianconi, 2013; Boccaletti et al., 2014; De Domenico et al., 2013; Kivela et al., 2014) to obtain information to describe and compare the signals.

The reason to use this approach instead of other ways to construct functional networks is because those usually require performing a pre-processing or symbolization, associated with loss of information (Kantz and Schreiber, 2004). Also, one of the advantages of using complex networks for analyzing time series is that they are becoming powerful tools when one seeks to construct feature vectors that can be used to automatically feed classifiers with low computational cost (Lacasa et al., 2015).

2. MATERIALS AND METHODS

2.1. Experimental data

The data of pollutant used for this analysis correspond to ozone and nitrogen oxide concentration values collected from 2010 to 2017, with a frequency of 10 minutes between each measure. Then, these data were separated into the different months in order to perform the analysis. The measurements were performed at the urban station located in Lepanto, Córdoba (37.53° N, 4.47° W). The cited station belongs to the regional network in charge of controlling the air quality in Andalusia, co-financed by the Consejería de Medioambiente (Regional Environmental Department) and the European Union. This station is located at 117 m of altitude and the average temperature and solar radiation is maximum on July and minimum on January every year. The region where it is placed is the western part of Andalusia (Spain). Since as exposed previously (Domínguez-López et al., 2014), this area meets the weather conditions (high temperatures and solar radiation), orographic (the valley of the Guadalquivir river) and anthropic ones to be potentially vulnerable to pollution by ground-level ozone and nitrogen oxide. The climate of the zone of study, according to the Köppen-Geiger classification, is defined as Csa, with warm average temperatures and hot and dry summer. Furthermore, the city of Córdoba is surrounded by two main industrial parks. One of them is located at southwest, and the other is at east. Moreover, there is a highway at southeast with frequently high traffic volume from both directions.

Authors have also employed the temperature, wind direction and average solar radiation in this work. They are shown further in the text (see Figure 7), were they are plotted along with the results in question, in order to illustrate their apparent relationship. These meteorological quantities have been provided by the Andalusian Agency of Energy.

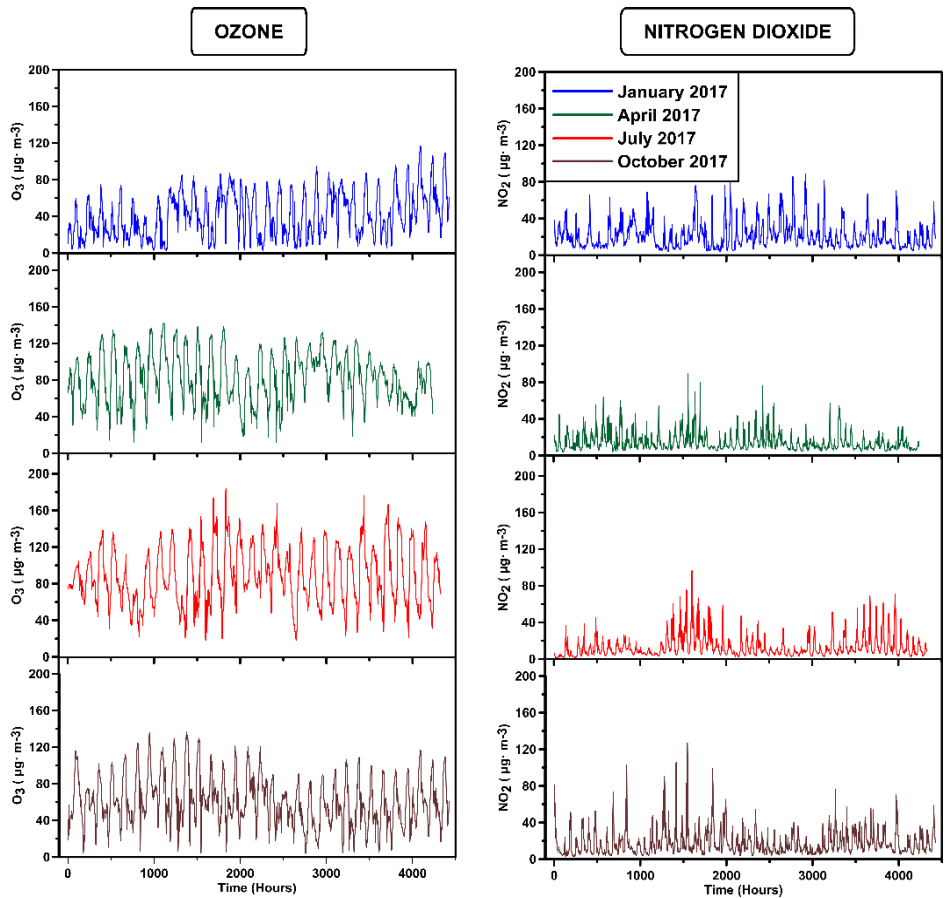

Figure 1: Sample time series of $O_3$ and $NO_2$ for four different months (year 2017).

2.2. Visibility graph

A graph can be defined as a set of vertices, points or nodes connected to each other by lines that are usually called *edges*. A tool to transform time series into a graph was presented in the last decade (Lacasa et al., 2008). This new complex network receives the name of *Visibility Graph* (VG) and has been proven to inherit many of the properties of the original signal (Lacasa and Toral, 2010). This means that, for instance, a periodic time series would lead to a regular graph and a fractal series to a scale-free one.

In order to construct the visibility matrix which contains the information of all the nodes in the system, it is necessary to stablish a criterion to discern whether two points would be connected or not. This criterion reads as follows: two arbitrary data from the time series $(t_a, y_a)$ and $(t_b, y_b)$ have visibility (and would become two connected nodes in the graph) if any other data point $(t_c, y_c)$ between them $(t_a < t_c < t_b)$ fulfills the following condition:

$$y_c < y_a + (y_b - y_a)\frac{t_c - t_a}{t_b - t_a} \tag{1}$$

The result of applying this visibility method is a *NxN* adjacency binary matrix, being *N* the number of points in the set. Each row of the matrix contains the information of a different node. For example, an element as $a_{ij} = 1$ means that the node $i$ and $j$ have visibility; whereas $a_{ij} = 0$ means that there is no edge between them. The resulting matrix has several properties that can be used to simplify the algorithm and thus reduce the computational required time: it is a hollow matrix ($a_{ii} = 0$), symmetric ($a_{ij} = a_{ji}$) and all the nearest neighbors have visibility between each other ($a_{ij} = 1$ for $j = i \pm 1$). Its general form is shown below:

$$A = \begin{pmatrix} 0 & 1 & \cdots & a_{1,N} \\ 1 & 0 & 1 & \vdots \\ \vdots & 1 & \ddots & 1 \\ a_{N,1} & \cdots & 1 & 0 \end{pmatrix} \tag{2}$$

In Figure 2, the application of the VG to two arbitrary time series is shown, highlighting the connections of two given time points (nodes in the graph) for the sake of clarity.

### 2.3. Degree centrality

To study the main properties of a complex network, centrality parameters become convenient mathematical tools to take into account. This kind of parameters measure the node importance within the graph in relation to the others by different approaches (Latora et al., 2017).

A very frequently used centrality parameter and an important feature of graphs in general is the degree. The degree of a node ($k_i$) can be defined as the number of nodes that have reciprocal visibility (in an undirected graph) with the first one ($k_i = \sum_j a_{ij}$). In Figure 2, the degree of the node that is highlighted is $k = 6$ for $O_3$ and $k = 3$ for $NO_2$.

From the degree of each one of the nodes present in the VG, it is possible to obtain the degree distribution of the sample $P(k)$, which is nothing but the probability that every degree has within the graph. This distribution can be analyzed to get a deep insight of the intrinsic nature of the time series, as previously demonstrated by (Lacasa et al., 2008; Mali et al., 2018; Pierini et al., 2012). The degree distribution of VGs whose right tails can be fitted to a power law in the way $P(k) \propto k^{-\gamma}$, are associated to fractal time series (Lacasa et al., 2008). The right tails are related to hubs, which are unlikely highly-connected nodes in the graph and therefore points with large values of degree. In a log-log plot, one can fit $P(k)$ to a simple linear regression, obtaining the so-called $\gamma$ coefficient, which has been directly related to the Hurst exponent of the time series in the Brownian motion (Lacasa et al., 2009).

2.4. Multiplex visibility graph

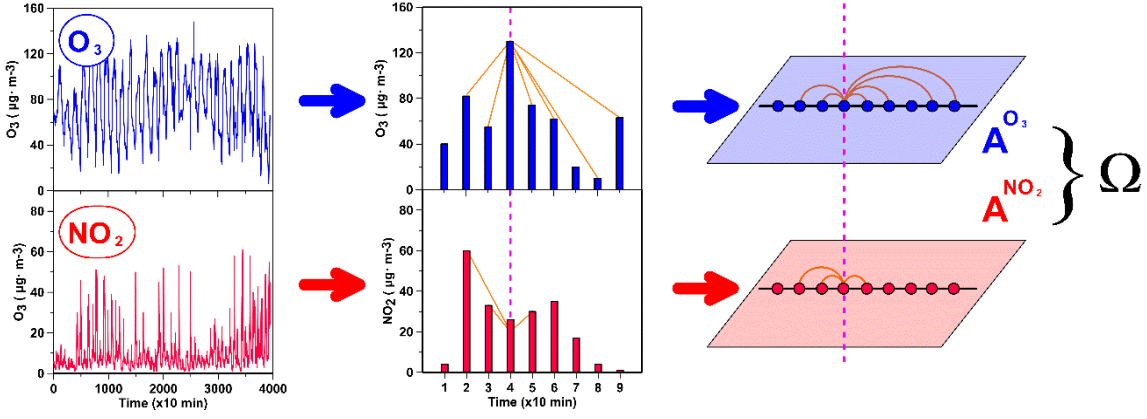

Figure 2: Time series of ozone and nitrogen dioxide concentrations (left) are transformed into complex networks through the VG algorithm (center), which is defined by an adjacency matrix ($A^{O_3}$ and $A^{NO_2}$). Finally, the two of them are combined to create a two-layered MVG, called Ω (right image).

In the case of a multivariate time series of $M$ variables, it is possible to construct a $M$-dimensional network from the VG of each one for the description of the signals (Lacasa et al., 2015). This multilayer network is called *Multiplex Visibility Graph* and each one of its $M$ layers corresponds to the VG of one of the variables from the underlying time series (see Figure 2). The MVG is represented by a vector of adjacency matrices $\Omega = \{A^{[1]}, A^{[2]}, ..., A^{[M]}\}$, being $A^{[\alpha]}$ the matrix corresponding to the VG of the $\alpha$-dimension (or layer in the multiplex) from the multivariate time series.

When it comes to analyzing the information that lies within these complex multilayer networks, there are several measures that can be used (Nicosia and Latora, 2015). Here, two quantities have been chosen. The first one is the *Average Edge Overlap (ω)*, that measures the number of layers on which a given edge between two nodes is found, on average. The other one, that captures the presence of inter-layer correlations of the degree distributions between two layers $\alpha$ and $\beta$, is the so-called *Interlayer Mutual Information* ($I_{\alpha,\beta}$). These layers in the presented study correspond to the VGs of $O_3$ and its precursor $NO_2$ concentration time series, hence the notation $I_{O_3,NO_2}$ will be used in this work.

The formula for the calculation of $\omega$ is presented in Equation 3, where $\delta_{0,\Sigma_\alpha a_{ij}{}^{[\alpha]}}$ corresponds to a Kronecker Delta and the other quantities are already defined.

$$\omega = \frac{\sum_i \sum_{j>i} \sum_\alpha a_{ij}{}^{[\alpha]}}{M \sum_i \sum_{j>i}(1 - \delta_{0,\Sigma_\alpha a_{ij}{}^{[\alpha]}})} \qquad (3)$$

The maximum possible value of this quantity is $\omega = 1$, and corresponds to the case where all the layers are identical. On the other hand, the minimum value is $\omega = 1/M$ (being $M$ the number of layers), meaning that a case with each edge in the multiplex existing just in one layer.

In Equation 4, $I_{\alpha,\beta}$ is defined:

$$I_{\alpha,\beta} = \sum_{k^{[\alpha]}} \sum_{k^{[\beta]}} P(k^{[\alpha]}, k^{[\beta]}) \log \frac{P(k^{[\alpha]}, k^{[\beta]})}{P(k^{[\alpha]})P(k^{[\beta]})} \qquad (4)$$

Where $P(k^{[\alpha]}, k^{[\beta]})$ is the joint probability of finding a node having a degree of $k^{[\alpha]}$ in the layer $\alpha$ and $k^{[\beta]}$ in the layer $\beta$. This joint probability is computed as follows:

$$P(k^{[\alpha]}, k^{[\beta]}) = \frac{N_{k^{[\alpha]},k^{[\beta]}}}{N} \qquad (5)$$

With $N_{k^{[\alpha]},k^{[\beta]}}$ being the number of nodes that have the corresponding degree of $k^{[\alpha]}$ and $k^{[\beta]}$ in layers $\alpha$ and $\beta$, respectively. Since $N$ is the total amount of nodes, it must be fulfilled that it is equal to the sum over all the possible $N_{k^{[\alpha]},k^{[\beta]}}$ values.

## 3. RESULTS AND DISCUSSION

### 3.1. Exploratory analysis

Before applying the MVG methodology, authors have performed a preliminary analysis of the data employed in this work. To do so, the first feature to consider has been the distribution of the degree of the independent VGs obtained from each time series. Some examples of these distributions can be regarded at Figure 3 (a and b). In these plots, only four months (the same ones for both) are depicted for illustrative purposes: January, April, July and October. The reason for choosing these months is that they are equally spaced through the year, each one represents a different season and they have been used in previous works in the same location (Carmona-Cabezas et al., 2019a; Jiménez-Hornero et al., 2010a). The year shown in this case is 2017, the most recent one that has been used here. All data used has not undergone any deseasonalizing because, as it has been previously discussed, VGs are not suitable for this kind of preprocessing approaches (Lange et al., 2018).

Looking at the mentioned figures, it can be noticed that both pollutants present a power-law behavior in the tail of the degree distribution obtained from their respective VGs, which points to fractal behavior of the time series. As commented in Section 0, from the linear regression in the log-log plot of this tail, one is able to obtain the $\gamma$ coefficient. At plain sight, it can be already seen Figure 3a and b how the distribution of the degree for ozone varies more along the year than those of the nitrogen dioxide in the given examples. Authors have computed the $\gamma$ coefficient of each month from 2010 to 2017 and shown their average monthly value in Figure 3c and d; where the previous statement is checked. The error bars correspond to the standard deviation of that quantity along the studied years. Therefore, the ozone concentration has a different behavior along the seasons, being the mean values of the coefficients between 2 and 4.25. Moreover, it can be appreciated that those are also irregular from one year to another, as the standard deviation is higher in comparison to the second gas by looking at Figure 3c and d. This was already seen in a previous work by the authors (Carmona-Cabezas et al., 2019a). On the other hand, $NO_2$ coefficients do not vary as much as the previous one, being always its value between 2.5 and 3.25. These results seem to indicate a different trend in the likeliness of hubs coming from VGs of $O_3$ and $NO_2$ and so distinct unlikely large values variation. Although in some cases the physical meaning of those coefficients has been related to fractal parameters such as the Hurst exponent (see Section 0 for more details), authors have employed these as a preliminary study before going deeper into the analysis with MVGs.

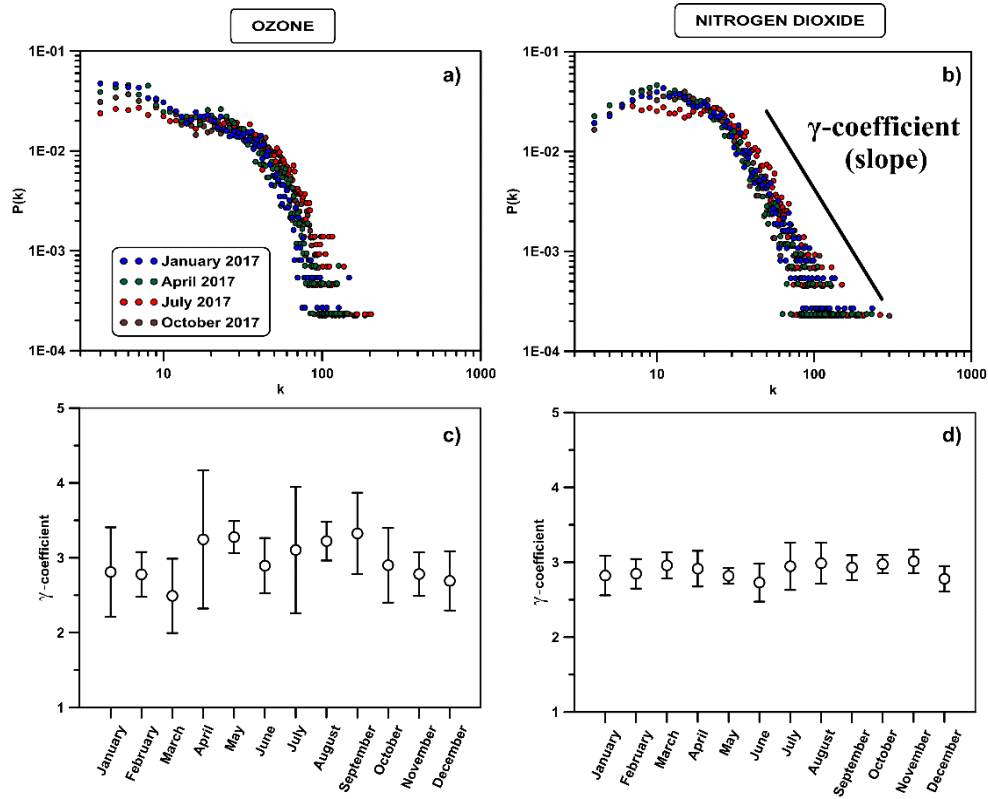

Figure 3: Degree distributions of four example months in 2017 for $O_3$ (a) and $NO_2$ (b) and the actual values of the γ coefficient obtained for every studied month and averaged from 2010 to 2017 (c and d). The error bars come from the standard deviation of the values obtained for all the years.

From the degree computed for each node of a given VG, it is possible to obtain also its total average value, which has been used in previous papers to describe the behavior of the maxima in the series (Carmona-Cabezas et al., 2019b; Donner and Donges, 2012). As in previous studies, a clear seasonal pattern is observed for both pollutants in Figure 4. The values of this figure correspond to the average obtained over the whole period from 2010 to 2017, while the error bars come from their standard deviation. This seasonality is less intense in the case of $NO_2$, which is in accordance to what was discussed in Figure 3.

In both cases, the maxima of the average degrees correspond to summer months. It was expected in ozone, since those months have the most favorable conditions for its creation and therefore there will be a higher number of maximal values (hubs) that increase the average degree. Nevertheless, the behavior of nitrogen dioxide is not as acute during summer. This difference might be due to the different factors that influence on both pollutants. Results point to the possibility that this quantity could be used to identify the known correlations between the two pollutants, and hence, authors have tested this hypothesis by using the MVG parameters. More precisely, $I_{\alpha,\beta}$ that is directly based on the degree of the two time series.

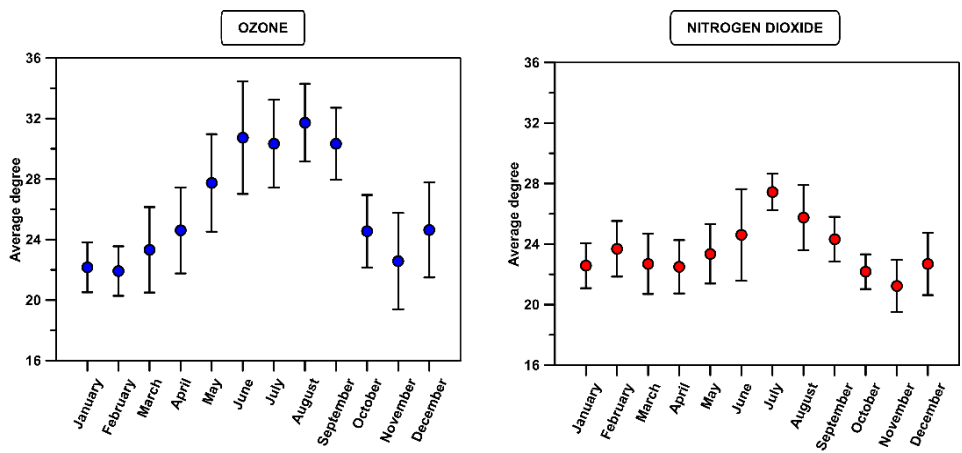

Figure 4: Monthly average degree values for each pollutant from 2010 to 2017. Again, the standard deviation along the different years is reflected through the error bars.

3.2. MVG analysis

After applying the VG algorithm to transform the $O_3$ and $NO_2$ concentration time series into complex networks, the MVGs for each month were built. With these multilayer networks, it was possible to compute $\omega$ and $I_{O_3,NO_2}$ for all the months considered. Figure 5a shows that $\omega$ values obtained are very similar for all the months and during the different studied years. This is clearly seen, as it has been averaged over the different studied years (2010-2017) and the standard deviation is as well remarkably low.

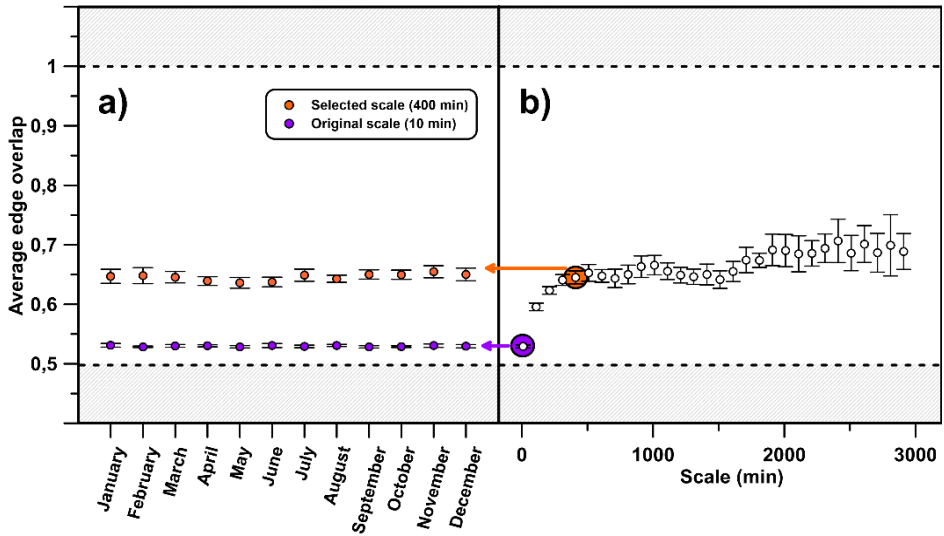

Figure 5: a) Parameter $\omega$ is computed for each month and then averaged over all the analyzed years (2010-2017), first with the original time scale (purple) and finally with a bigger one (orange) derived from the analysis shown in b). In this one, $\omega$ is recalculated with several scales for constructing the MVGs. The selected scale is taken from the point where the saturation starts. The white area between the grey zones corresponds to all the possible values that $\omega$ can have for MVGs with two layers.

As $\omega$ accounts for the overall coherence of the original time series, high values of it indicates a strong correlation in the microscopic structure and vice versa (Lacasa et al., 2015). In the context of VGs applied to time series, the microscopic structure can be understood as the most likely scale of the edges of the graph. This scale can be considered as the time resolution of the recorded signal. For that reason, a study was performed to see how the $\omega$ parameter behaves when the minimal size of the edges of the concentration of $O_3$ and $NO_2$ is changed (see Figure 5b).

The result is that the edge overlap increases rapidly as the number of points in the sample is decreased (the VGs become smaller and their edges bigger) and then it starts to saturate around $\omega \approx 0.65$ for scales greater than 400 minutes per measured point (about 6 hours). On the other hand, for the original scale of the measure $\omega \approx 0.53$, which is very close to the minimum value that quantity can hold ($\omega = 0.5$). From that, this result can be understood as follows: the correlation between the $O_3$ and the $NO_2$ concentrations is very weak at a scale of 10 min (the order of magnitude of the smallest edges in the graphs). Nevertheless, this correlation becomes greater as the minimal size of these edges (the microscopic scale) is increased, until this rising saturates. Authors attribute this effect to the time needed for the two systems (gases) to reach coupling, so that it affects the correlation between their concentrations. The posterior saturation would correspond to the fact that they are already coupled for higher scales, so there is no increase on $\omega$ due to physical processes. The slight increase observed would be due to the reduction of the number of nodes of the VGs (mathematical artifact). Hence, this approach to determine the time scale at which the saturation starts could be used to determine the effective time of a given reaction. This could be useful for later works describing the relationship between some other pollutants.

As a previous step before introducing the computed $I_{O_3,NO_2}$, authors have depicted in Figure 6 the quantity $P(k^{O3}, k^{NO2})$. Again, the same months as before are used in these plots as an example, for the sake of clarity. Also, only one year is taken (2017, the most recent), since the results have been found to be equivalent and the same conclusions could be drawn for different years. The meaning of these figures can be interpreted as combined degree distributions, where the colors indicate the probability of the two VGs having degree $k^{O3}$ and $k^{NO2}$, simultaneously. It is regarded how the most likely combinations of values of k are those of the lowest values of the degree ($k \in [0, 50]$). By contrast, as the degree increases, the joint probability becomes less and less significative.

It must be pointed out that the probability approaches asymptotically to both X and Y axis. It means that as the degrees increase, the probability of encountering relatively similar both $k^{O3}$ and $k^{NO2}$, decreases exponentially. This translates into the alternation between the hubs of the two time series. The reason behind this is the time shift that exists between both NO$_2$ and O$_3$ maximal concentrations throughout the day (previously mentioned). One of the reactions that governs the ozone creation and destruction is $NO_2 + O_2 \leftrightarrow O_3 + NO$ (Graedel and Crutzen, 1993). According to this one, when the ozone reaches a maximum, the concentration of nitrogen dioxide decreases in general, leading to what has been argued here.

Another characteristic of the plot is that as the year advances, the distribution of the joint probability changes, being more concentrated around the $k^{NO2}$ axis for January, while in July it is more equally distributed (April and October present an intermediate behavior). Since the value of the degree and concentration are related (Carmona-Cabezas et al., 2019b; Pierini et al., 2012), the interpretation can be seen as follow:

- In January, $P(k^{O3}, k^{NO2})$ is more concentrated to specific combinations of degrees of the gases, specially, low $k^{O3}$ and higher $k^{NO2}$. This translates into fewer values of the temporal series that are considerably correlated and thus, the overall correlation will decrease (Figure 7). As a result, high concentrations of NO$_2$ will not necessarily lead to greater production of ozone, as can be regarded in Figure 1; where there are many days with extreme values of nitrogen dioxide concentration, whereas the ozone levels remain at minima with respect to the rest of the year. The reason is that although high concentrations of $NO_2$ are available, there is not enough solar radiation to make the optimal interaction possible.

- In July, the concentration of ozone rises, as it widely known. In Figure 6, $P(k^{O3}, k^{NO2})$ is in this case more homogeneous and non-null for the values of k where the vast majority of points are located: $k \in [0, 50]$. The result of this will be an increase in the $I_{O_3,NO_2}$ that is shown on average in Figure 7. In contrast to January, now the extreme concentrations of $NO_2$ coincide with those of $O_3$, for instance in Figure 1 for July, from $1.5 \cdot 10^4$ to $2 \cdot 10^4$ minutes both reach their highest concentrations (taking into account that there exist a time delay between both quantities).

- Lastly, April and October are intermediate cases, where the distribution of $P(k^{O3}, k^{NO2})$ is neither as acute as in January, nor as regular as July. In both cases, the probability is higher for finding low $k^{O3}$ when $k^{NO2}$ is greater.

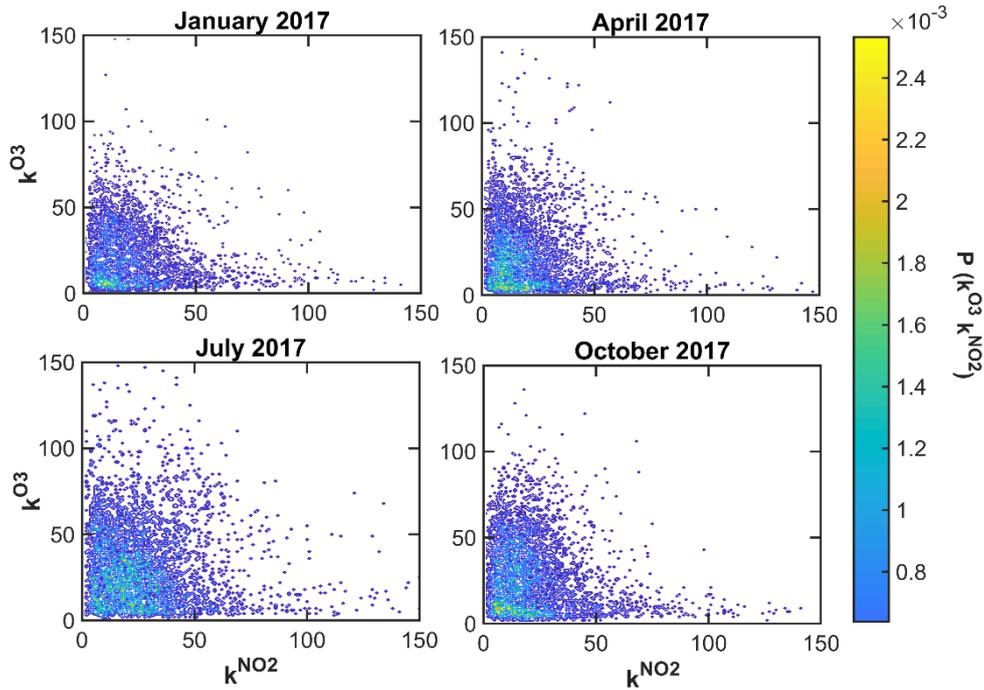

Figure 6: Graphical representation of the joint probability distribution of the degrees of both layers for four sample months in 2017, the most recent year. Each point represents the probability that degree is exactly $k^{O3}$ in the ozone VG, while is $k^{NO2}$ in the nitrogen dioxide case, at the same time node.

The monthly computed values of $I_{O_3,NO_2}$ are presented in Figure 7a, where the different years are indicated by several colors. Figure 7b shows the monthly average value (over all the mentioned years) and the standard deviation along with the temperature, the average global radiation (normalized for the sake of clarity) and wind direction.

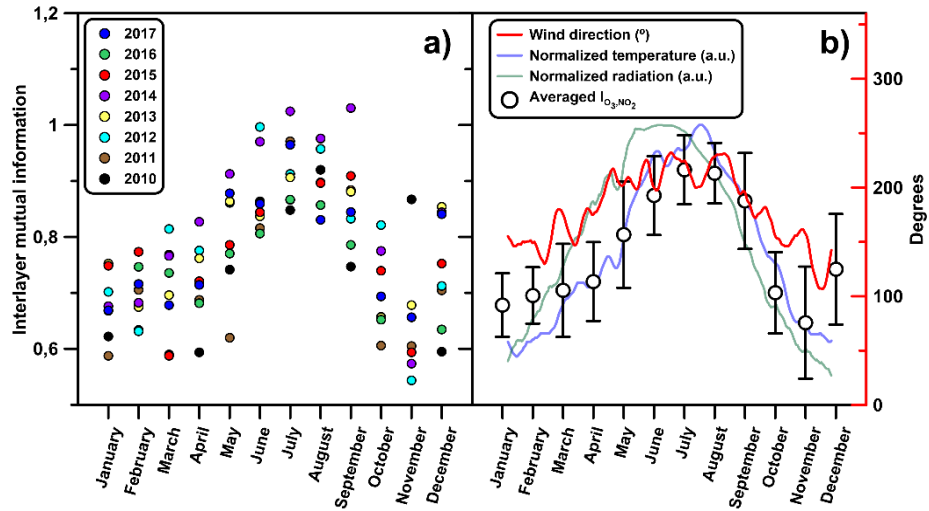

Figure 7: a) Seasonal pattern obtained for $I_{O_3,NO_2}$ for each one of the studied months and years. b) Averaged $I_{O_3,NO_2}$ over the whole period (2010-2017) for every month and normalized temperature, solar radiation and wind direction for comparison. 0, 90, 180 and 240 degrees correspond to winds coming from the East, North, West and South directions, respectively.

It can be appreciated that there is a sensible increase of the mutual information between $O_3$ and $NO_2$ from May to September, coinciding with the period of highest temperature and radiation. By definition, $I_{\alpha,\beta}$ describes the average correlation of the degree distributions and these can be used to describe the nature of the time series (Lacasa et al., 2008; Mali et al., 2018). From that, it could be inferred that the greater this parameter is, the more similarly are expected to behave the signals underlying the VGs studied. Therefore, in this case the degree distributions of the two pollutants are more correlated during the summer months and the opposite during winter. It might indicate that they will have a more similar behavior in the first case. For instance, $O_3$ is known to have multifractal nature (Carmona-Cabezas et al., 2019a; He et al., 2017; Jiménez-Hornero et al., 2010a; Pavon-Dominguez et al., 2013), hence $NO_2$ could be expected to have higher multifractality in the summer months, being in accordance with the findings reported in a prior work for the same area (Jiménez-

Hornero et al., 2010b). A higher multifractality is understood as a greater degree of multifractal behavior, which is usually related to complex systems with fluctuations appearing at different scales (Kantelhardt, 2011).

The reason for this clear seasonal pattern seems obvious looking at the distribution of solar radiation and temperature, which change greatly along the seasons in the study area (both maximal in summer and vice versa in winter). The first one of them (solar radiation) is the main source of energy for the photochemical reaction needed for the formation of O3 in the troposphere (Graedel and Crutzen, 1993), since nitrogen dioxide ($NO_2$) and oxygen ($O_2$) are recombined and produce ozone gas in the presence of ultraviolet light. The second one has been found to have relation as well with the production of ozone from nitrogen oxides in previous works in the same area (Pavón-Domínguez et al., 2015). Many of the previous studies are based on multifractal approaches to analyze time series, which have some disadvantages such as the need for choosing a scale interval where the searched behavior holds, leading to errors in the computation and some undesired ambiguities of the results. On the other hand, VGs are univocal for a given time series and their computation do not have any error associated, since their results are based on basic graph theory arithmetic.

Furthermore, regarding the average wind direction that can be seen in Figure 7b, it is clearly seen how the direction is roughly ~220° when the correlation between $NO_2$ and $O_3$ reaches it maximal value for summer, as commented before. This corresponds to wind coming as an average from the South-West. This fact has been previously described for the Guadalquivir Valley (Guardans and Palomino, 1995), where the pressure and temperature differences make more predominant the wind coming from South-West direction when the temperature is maximum (summer). The reason behind this phenomenon is that the air mass moves from the plain areas towards the upper parts of the valley (North-East). The opposite case occurs in winter following an equivalent reasoning.

It must be pointed out the fact that in that predominant direction, the main sources of $NO_2$ in the vicinities of Cordoba are located, since there are two of the main industrial parks of the city. Also, the most populated capital of the region (Sevilla) is situated in that direction, as well as the main highway that connect the two cities, which is one of the most transited roadways in Spain. This fact further corroborates the adequation of the correlation results, as was seen with the temperature and solar radiation previously.

4. <u>CONCLUSION</u>

All the stated results confirm the capability of the two parameters provided by the MVGs for describing the interaction between ozone and nitrogen dioxide in the troposphere. On the one hand, authors consider that the first studied parameter ($\omega$) may be used to infer the time shift in the coupling of the two systems, represented by the two layers. Given that the value of this parameter does not vary significatively neither in the different seasons, nor along the years, it means that $\omega$ does not depend on any external factors, such as meteorological features. Authors believe that this quantity might be used to check theoretical models of $O_3$ - $NO_2$ interaction, although further investigation will be needed.

On the other hand, $I_{O_3,NO_2}$ and $P(k^{O3}, k^{NO2})$ could be used to have an insight of the correlation in the behavior of the series (from the degree, which is associated to the concentration itself). Regarding the first one, one can obtain an overall look of how the $NO_2$ and $O_3$ are correlated depending on the concentration. For instance, it is possible to see how great values of $NO_2$ correlate with large or low ones of $O_3$, and vice versa. Furthermore, the second quantity corresponds to a numerical value that measures the correlation of the whole set of series that are transformed into the different layers of the MVG.

To authors' mind, the outcomes of this research support the capability of multilayer complex networks for looking at the relation between several variables such as atmospheric pollutants. By using this approach, one can take advantage of some of its assets, as the computation efficiency or the univocity of the results. All this situates MVG as a suitable complementary technique for tackling analyses within the environmental problem. It as well compatible with others that have demonstrate to give satisfactory results, such as multifractal algorithms, as it has been demonstrated (Carmona-Cabezas et al., 2019a; Mali et al., 2018; P. Pavón-Domínguez et al., 2017). Hence one possible application of MVGs in this context could be the enlargement of databases used for predictive techniques that rely on data mining and machine learning. For this aim, it remains open the study of more pollutants and other variables through the methodology analyzed here. Since it is possible to focus MVGs on only two variables or to see the correlation of many of them at the same time, it allows a flexible analysis. All these considerations could come in handy for describing the many mechanisms involved in the dynamics of the photochemical smog for future works.

## 5. ACKNOWLEDGEMENTS

The FLAE approach for the sequence of authors is applied in this work. Authors gratefully acknowledge the support of the XXIII research program (2018) of the University of Cordoba.

## **Declaration of interest**

There is no conflict of interests to declare.

**HIGHLIGHTS**

- Multiplex visibility graphs are used to check correlations between $NO_2$ and $O_3$.

- $NO_2$ and $O_3$ have different behavior of their degree distributions along the year.

- Average edge overlap between the two pollutants remains constant in every case.

- Interlayer mutual information evolves with a seasonal behavior every year.

Graphical Abstract

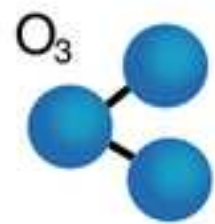
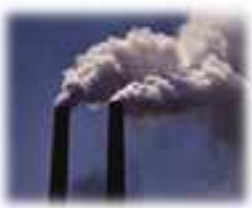
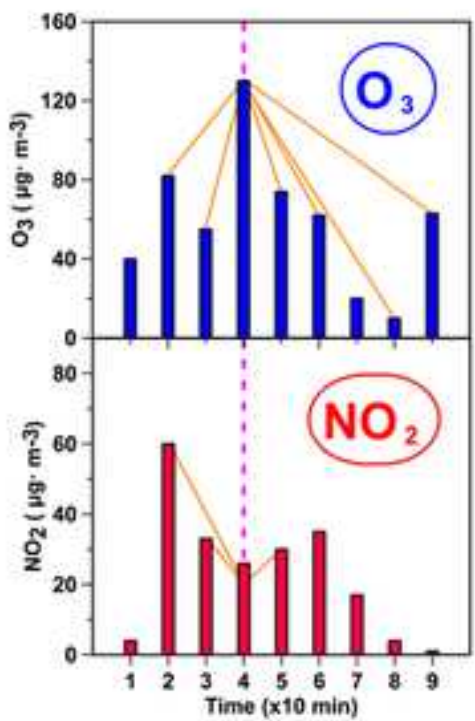
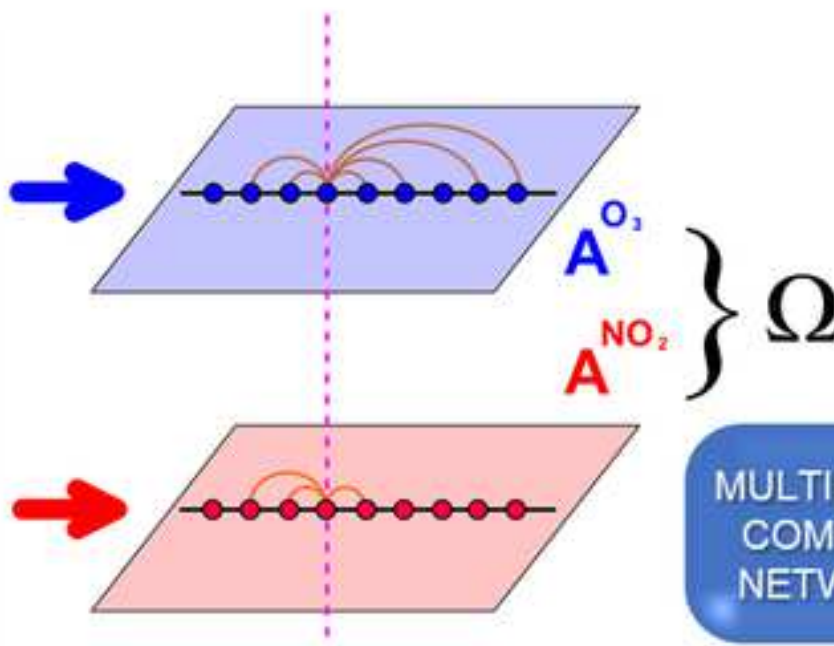
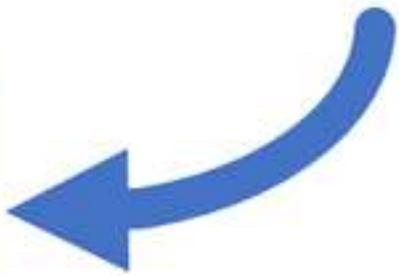